\documentclass{sf2a-conf}
\usepackage{graphicx}

\begin{document}
\TitreGlobal{SF2A 2008}
\title{The Gaia satellite: a tool for Emission Line Stars and Hot Stars}
\author{Martayan, C.$^{1,}$} \address{Royal Observatory of Belgium, 3 avenue circulaire 1180 Brussels, Belgium}
\address{GEPI, Observatoire de Paris, CNRS, Universit\'e Paris Diderot; 5 place Jules Janssen 92195 Meudon Cedex, France}
\author{Fr\'emat, Y.$^1$} 
\author{Blomme, R.$^1$} 
\author{Jonckheere, A.$^1$} 
\author{Borges, M.} \address{Fizeau, Observatoire de la C\^ote d'Azur, France}
\author{de Batz, B.$^2$} 
\author{Leroy, B.} \address{LESIA, Observatoire de Paris, 5 place Jules Janssen 92195 Meudon Cedex, France}
\author{Sordo, R.} \address{Padova Astronomical Observatory, Italy}
\author{Bouret, J.-C.} \address{Laboratoire d'Astrophysique de Marseille, France}
\author{Martins, F.} \address{GRAAL-Universit\'e Montpellier 2, France}
\author{Zorec, J.} \address{Institut d'Astrophysique de Paris, France}
\author{Neiner, C.$^2$} 
\author{Naz\'e, Y.} \address{Institut d'Astrophysique de Li\`ege, Belgium}
\author{Alecian, E.} \address{RMC Kingston, Canada}
\author{Floquet, M.$^2$} 
\author{Hubert, A.-M.$^2$} 
\author{Briot, D.$^2$} 
\author{Miroshnichenko, A.} \address{University of North Carolina at Greensboro, USA}
\author{Kolka, I.} \address{Tartu Observatory, Estonia}
\author{Stee, P.} \address{Fizeau, Observatoire de la C\^ote d'Azur, France}
\author{Lanz, T.} \address{Dpt of Astronomy, University of Maryland, USA}
\author{Meynet, G.} \address{Observatoire de Gen\`eve, Switzerland}
\runningtitle{ELS and HS with Gaia}
\setcounter{page}{237} 

\index{Martayan, C.}
\index{Fr\'emat, Y.} 
\index{Blomme, R.} 
\index{Jonckheere, A.} 
\index{Borges, M.} 
\index{de Batz, B.} 
\index{Leroy, B.}
\index{Sordo, R.} 
\index{Bouret, J.-C.} 
\index{Martins, F.} 
\index{Zorec, J.} 
\index{Neiner, C.} 
\index{Naz\'e, Y.} 
\index{Alecian, E.} 
\index{Floquet, M.} 
\index{Hubert, A.-M.} 
\index{Briot, D.} 
\index{Miroshnichenko, A.} 
\index{Kolka, I.} 
\index{Stee, P.} 
\index{Lanz, T.} 
\index{Meynet, G.} 

\maketitle
\begin{abstract}
The Gaia satellite will be launched at the end of 2011. It will observe at least 1 billion
stars, and among them several million emission line stars and hot stars.
Gaia will provide parallaxes for each star and spectra for stars till V magnitude equal to 17.
After a general description of Gaia, we present the codes and methods, which are currently developed by our team. 
They will provide automatically the astrophysical parameters 
and spectral classification for the hot and emission line stars in the Milky Way and other
close local group galaxies such as the Magellanic Clouds.
\end{abstract}
%
\section{Introduction: The Gaia space mission}

The Gaia space mission will be launched in 2011/2012. It will orbit at the anti-solar L2 point. Its lifetime is
expected to be 5 years. Onboard, there are 3 different instruments:  ASTRO, which will provide astrometric
measurements (parallaxes, proper-motions) for all stars down to V magnitude 20 with an accuracy 200 times better
than Hipparcos. There are 2 spectrophotometers BP/RP (R$\sim$100, 320--660nm and 650--1000nm) and a low/medium
resolution spectrograph called the Radial Velocity Spectrometer (R=5000 to 11500, 847-874nm, designed for GK stars).
This last one will provide spectra for stars till V magnitude equal to 17, and will be used to determine the radial
velocity of the stars (see Viala et al. 2008). It is expected that 1 billion of stars will be observed by Gaia, on
average 70 times in 5 years (but only 40 times for the RVS). Among them, using the IMF from Kroupa (2001),  it is
estimated that there are, at maximum, 68 million of hot stars (HS), and 6 million of emission line stars (ELS).  Due
to the huge quantity of data (200 teraBytes by year), all the data-reduction and the scientific analysis must  be
performed automatically via new software based  on java programming language.

\section{Introduction: Hot stars and Emission Line Stars}
The hot stars are defined as stars with effective temperature higher than 7500K. They concern main sequence OBA stars
but also Supergiant O stars, etc. The emission line stars are the stars with emission lines in their spectra (in the RVS
domain but also in others parts of the spectrum). From time to time, some ELS exhibit emission lines in the UV-Visible part
of the spectrum but not in the RVS domain, in which CaII triplet and/or Paschen lines are present. This is the case for
example of the late-type Be stars. More generally, ELS can be found everywhere in the HR diagram, they range from hot to
cool stars and from young to evolved stars. As examples, one can cite the WR, LBV, Oe, Of, Be, supergiant Oe to Ke, PNe,
HBe/Ae, B[e], Mira Ceti e, TTauri, UV Ceti, Flare stars, etc.

With Gaia, we shall obtain for these stars:
\begin{itemize}
\item with ASTRO: the proper motions and distances, 
\item with the BP/RP spectrophotometry: the ELS detection (for example in H$\alpha$), the stellar photometric
classification, and photometric fundamental parameters.
\item With the RVS spectrograph: the ELS detection, the radial velocity, the spectroscopic fundamental parameters,
the stellar spectral classification.
\item An indice of the spectroscopic and photometric variability of the stars because they will be observed several times
during the 5 years of the mission.
\item Indications about potential behaviour differences between HS and ELS from our Galaxy and close local group galaxies,
because the bright stars of these galaxies, which correspond mainly to hot stars, will be observed by Gaia.
\end{itemize}

Combining these informations, we shall determine:  
\begin{itemize}
\item 3D static and dynamic maps. These 2 maps allow to study the open cluster membership, the site of origin of the stars,
the site of formation, and then to characterize the stars.
\item The fundamental parameters and spectral classification of stars and where possible the parameters of the disks.
\item Statistical links between the lines (amount of emission between H$\alpha$ and Pa for example). 
\item With the distance, one can provide statistical relation between Mv and spectral types and find potential
outliers among OB stars.  This deviation could be interpreted in term of fast rotation effects (see Lamers et
al. 1997).
\item The interstellar reddening of the stars to correct their photometry. The remaining reddening will be
interpreted as due to circumstellar matter/disk (Be, HBe/Ae stars).
\item Previously unknown ELS (Be, WR), in case of Be stars, the Be phenomenon is transient and a Be star could be seen like a B
star at one epoch and several years after seen like a Be star (due to the disk variability).
\item The evolutionary status of Be stars. It will be possible to compare the
status determined spectroscopically and photometrically via derredened abolute magnitudes.
\item An index on the deviation from the expected behaviour of WR stars.
\item The correct classification of B[e] stars. 50\% of them are currently badly or not classified: SgB[e], PN[e],
HB[e] because their distance is not yet known.
\item The percentage of binarity. This is an important issue to understand the behaviour of some stars: what is the
rate of binaries among Be stars? 75\% according to McSwain \& Gies (2005) or 30\% according to Porter \& Rivinius
(2003)? What is the rate of binaries among hot stars? 60\% for O stars according to Sana (2008), 30\%  
according to Porter \& Rivinius (2003) for B stars.
\end{itemize}

\section{Algorithms and methods}
In this section, we briefly present the methods of algorithms currently developed in order to obtain the astrophysical
parameters of the stars and reach the goals enumerated above. They are elaborated in Coordination Unit 6, 
in the Coarse Characterization of Sources scientific module (managed by C. Martayan), in Coordination Unit 8,
in the Extended Stellar Parametrizer scientific module (managed by Y. Fr\'emat), in the Hot Stars scientific module
(managed by R. Blomme), and in the Emission Line Stars scientific module (managed by C. Martayan).

\subsection{Photometric detection of ELS and pre-classification of HS/ELS}
The first possibility to detect the ELS and to pre-classify the HS is to use photometric colour-colour or
colour-magnitude diagrams. To do that, with the BP/RP spectrophotometry, it is possible to obtain magnitudes in
different filters similar to the classical Johnson or Geneva filters and draw the corresponding diagrams. Among
them, to detect the ELS, the CMD (R-H$\alpha$ vs. V-I) could be very useful as shown by Keller et al. (1999). Note
that due to the low resolution, only stars with strong emission in H$\alpha$ could be detected.

It is also possible to do the same kind of study with the RVS. We computed magnitude-filters in the RVS, defined in
areas with CaII lines or Pa lines or both, and in the continuum. The magnitudes are then normalized to the size of
the filters. Theoretical magnitudes are computed in filters where there are Pa+Ca lines (filter PaCa) and only one
single Pa line (filter PaS that corresponds to the Pa14 line, which is the alone Pa line not blended with CaII lines
in the RVS domain). Then the difference between the theoretical magnitude and the observed magnitude  in the 2
filters are obtained (PaCa$_{th}$-PaCa$_{obs}$ and PaS$_{th}$-PaS$_{obs}$). The difference between 2 filters in the
continuum is also determined in order to have an index on the slope of the spectrum and split the stars by
categories (hot/cool stars). The first tests on simulated spectra (from CU2 team) for Be, WR stars and normal stars
(from O to M stars) show that with the PaCa diagram, the ELS with strong or medium emission in the RVS are
detected (above a threshold defined by the upper limit of the normal stars). We also used observed spectra both for
ELS and HS  from various ground-based spectrographs, which are correctly pre-classified with these diagrams (ELS
or absorption). With the PaS diagram, we pre-classify the stars in spectral-types because the filter PaS allows to
know whether the star contains or not Pa lines and then whether the star is a hot star or not).

\subsection{Fundamental parameters determination}
The fundamental parameters determination for hot stars is performed using grids of NLTE models with winds. 
The observed spectrum is fitted with theoretical ones using the Simplex downhill method
(Nelder \& Mead 1965). First tests based on 1089 synthetic spectra with noise added, Teff, logg, Vsini randomly
chosen were performed. They indicate that in case of B stars, 61\% of Teff are determined with an error bar of 10\%,
65\% of logg are determined with an error-bar of $\pm$0.25dex ($\pm$6\%), 66\% of Vsini are determined with an
error-bar of $\pm$50\%. In case of O stars, the error-bars are greater than for B stars, because the lines in the
RVS spectra are weak. Due to its design, for HS the RVS displays only hydrogen lines (Pa lines) and from time to
time weak HeII lines,  which implies a poor precision on the Vsini. Full results are detailed in Fr\'emat et al.
(2007, 2008 in preparation),  and Martayan et al. (2008). Moreover, the parameters derived for the HS will be useful
for improving the galaxy models, in which there is a lack of HS population.
  
The same method of minimization/determination could be used to fit the emission lines with theoretical models  to
obtain parameters for the stellar winds and/or circumstellar matter/disks. Another technique for classical Be stars
can use the relation between the strength of emission in H$\alpha$ and Pa lines.  However, it is necessary to remove
the amount of emission due to CaII lines in the Ca--Pa blends. Using data from Briot (1981), we obtained an
interpolated distribution of cleaned Pa emission in the RVS domain using the Pa14 line as beginning point.

\subsection{Spectral classification}
There are different possibilities to spectroscopically classify the stars. The first one is to use the Teff-logg
plane and calibrations, which link fundamental parameters to spectral types (Zorec 1986, Bouret et al. 2003). The
second one is to use the equivalent widths of the lines and calibrations from Andrillat et al. (1995), and
Carquillat et al. (1997). However, to do that, we have to detect the presence of a line, to determine the parameters
of the lines (the wavelength, the intensity in emission or absorption, the equivalent width). For detecting the
lines, we developed different algorithms based on local signal to noise ratio variations, the local slope 
variation, and by gaussian fitting (see Viala et al. 2008, Fr\'emat et al. 2007). In case of high signal to noise
ratio (better than 20), the different methods provide good lines detection. In case of bad signal to noise ratio
(lower than 20), from time to time wrong lines are detected. The automatic identification is currently tested. It is
based on a cross-correlation of the theoretical and observed wavelength -differences between consecutive lines. The
EWs, FWHM, I, are then determined and added to the tables of observed lines and identified lines. 
Details will be given by Martayan et al. (in preparation).

\subsection{Classification with neural networks approach}
We explore also other kind of classication methods based on the neural network approach or on the decision tree
via the WEKA library. These algorithms need training samples of spectra for each kind of stars in order to teach the
neural networks. Then, they are tested with tests samples in order to determine the reliability of their automatic
classification. Finally, after having defined the rates of good/bad classification, they are used in the true data.
Currently, the first tests based on magnitude in filters (defined above) provide an excellent classification of stars,
but the 1000 spectra used here are synthetic noise-free ones and we need more true observed spectra for each kind of object
(ELS, HS) to correctly supervise the learning of the neural networks.

An important issue for these methods as well as for the other algorithms currently developed is to obtain in the near
future enough observed spectra to test and improve them before the launch of Gaia. In addition, it will be useful for
a more complete scientific analysis of the stars to have a dedicated telescope/instrument (probably a multi-objects
spectrograph) for the follow-up of the Gaia targets.

\section{Conclusion}
The Gaia satellite is one of the most ambicious space mission of the next decades. It will provide 
some astrophysical parameters for 1 billion of stars and among them for several million of hot stars 
and emission line stars both in the Milky Way and close local group galaxies. They will allow to 
elaborate 3D static and dynamic maps, statistical studies, giving new clues about the behaviour of the 
stars (origin, evolution, variability, binarity).
Different methods and algorithms are currently developed by our team in order to detect and identify the lines, 
to determine the fundamental parameters of the stars, and also to provide their spectral classification.


\begin{acknowledgements}
C.M. acknowledges funding from the ESA/Belgian Federal Science Policy in the 
framework of the PRODEX program (C90290).
C.M. thanks the SOC of the Gaia session for the invitation to present a talk 
and the AS-Gaia France for the financial support.
\end{acknowledgements}

\end{document}